\documentclass[12pt]{article}
\usepackage{hyperref}
\usepackage[english]{babel}
\usepackage{amsmath}
\usepackage{latexsym}
\usepackage{amssymb}
\usepackage[all]{xy}
\usepackage[dvips]{graphicx}
\usepackage{epsfig}
\usepackage{psfrag}

\numberwithin{equation}{section}
\numberwithin{table}{section}
\numberwithin{figure}{section}

\begin{document}

\begin{titlepage}

\begin{center}

\hfill  hep-th/0709.0308  \\
\hfill ITFA-2007-40

\vskip 0.5 cm {\Large \bf More Rings to rule them all : \\  \vskip 0.5cm  \Large \bf Fragmentation,  4D$\leftrightarrow$5D  and Split-Spectral Flows   }

\vskip 1.25 cm {   Xerxes D. Arsiwalla  \footnote{e-mail:  xarsiwal@science.uva.nl }    }\\
{\vskip 0.5cm  Institute for Theoretical Physics\\ University of Amsterdam\\
Valckenierstraat 65\\
1018 XE, Amsterdam\\
The Netherlands\\}

\end{center}

\vskip 2 cm

\begin{abstract}
\baselineskip=18pt
 In this note we set-up an explicit 5D construction of AdS-fragmentation, whereby a single black ring splits-up into a multi-black ring configuration.  Furthermore it is seen that these fragmented rings are equivalent  to a direct 5D lift of 4D multi-center black holes.  Along the way we also determine the  4D/5D  transformations relevant for  multi-center charges.  It is seen that the physical charges involved in black ring fragmentation are  Page charges arising due to 5D Chern-Simons terms. As an  application of these methods, we  reproduce the total angular momentum of concentric black rings, originally due to  Gauntlett and Gutowski.  Finally we provide a geometric interpretation of  fragmented black rings using the idea of split-spectral flows, which seeks to study charge shifts of a given black ring due to fluxes generated in a multi-ring background.  
\end{abstract} 
\setcounter{page}{0}
\end{titlepage}

\pagestyle{plain}
\baselineskip=19pt

\tableofcontents

\section{Introduction}                           
Starting from ( not very ) recent work in  \cite{GSY} and  \cite{GSY2}, a considerable interest has been generated in understanding   5D BPS degeneracies by constructing dualities to the better understood 4D sector \cite{BKW}, \cite{BK}, \cite{CGMS}, \cite{GHLS}. Matter of fact, this 4D/5D relation was put forth by  \cite{GSY} as a 5D version of the OSV conjecture \cite{OSV}
\begin{eqnarray}
Z_{BH}^{5D} =  Z_{BH}^{4D} = |Z_{top}|^2
\label{1.2}
\end{eqnarray}
Evidence for this proposal was sought for by matching the entropy of the  5D       BMPV  \cite{BMPV}  black hole in Taub-NUT space, to the entropy of a 4D Calabi-Yau black hole while making use of the M-theory $\leftrightarrow$ Type II A correspondence.  Moreover,  since $Z_{BH}^{4D}$ counts  degeneracies of single as well as multi-center black holes,  it was pointed out by  \cite{GSY2}  that   $Z_{BH}^{5D}$ must also account for equivalent multi-black objects in 5D, assuming  eq.(\ref{1.2}) holds. While a single-center BPS black hole  in 4D   just  lifts  to a 5D BMPV black hole; in \cite{GSY2} a rather interesting  result was demonstrated :  a 4D two-center charge configuration consisting of a D6 charged point-particle at the origin ( of ${\mathbb R}^3$ ) and a D4-D2-D0 charge  at a distance $|\vec{L}|$ from it, will  in fact lift to a supersymmetric black ring in 5D  Taub-NUT space. $|\vec{L}|$  now becomes a modulus on Taub-NUT denoting the distance of the ring from the origin. 

On a rather different footing, yet another offshoot of the OSV bandwagon was the work of  \cite{DGOV},  conceiving  baby universes as  finite ( but still relatively large  ) $N$  non-perturbative corrections  to the OSV conjecture.  These corrections go like $e^{-N}$ and are realised as instanton effects in the holographically dual gauge theory. In turn, the holomorphic sector of the gauge theory is dual to the topological string partition sum  $Z_{top}$.  The gravitational realisation of these corrections were proposed as  4D multi-center black hole  configurations,  which can be  generated via the mechanism of  AdS fragmentation  \cite{Br}, \cite{MMS}  of a single black hole at $x_0 \in {\mathbb R}^3$ into multiple  black holes at $\{ x_i  \in {\mathbb R}^3 \}$.  These multi-AdS throats  are associated to a gravitational  instanton action  which describes the amplitude for  tunneling, in Euclidean time, of a single  black hole to multi-black holes. Based on that, \cite{DGOV} forward the idea of a third quantized Hilbert space of baby universes. 

One of the motivations driving this note was to reconcile the two aforementioned  streams of thought. We  try and address some questions regarding  the fragmentation of black rings in 5D.  Analogous to the 4D case, where we saw how to split   D4-D2-D0 charges, here we start with a black ring in Taub-NUT, since this is   the pertinent  5D lift of a D4-D2-D0 black hole placed at a distance $|\vec{L}|$ from a single D6 charge ( the sole D6 here does not participate in fragmentation  ).  We then set up a  fragmentation ansatz for this single ring and see that it  splits up into non-concentric multiple black rings ( in general ).  This construction is subject to charge splitting constraints, which as we shall soon see will turn out to be  more subtle in the 5D case that they were in 4D due to the presence of cross-terms between multiple centers that must now be carefully   tendered.    

     On the other hand, one might fairly well ask whether the fragmented multi-rings constructed in this manner could as well have been obtained from a direct 5D lift of the 4D multi-center solution.   The answer turns out to be in the affirmative; and to do so we shall first reqiure to construct the  4D/5D dictionary for   multi-center charges.  Compared to  the  4D/5D map of  \cite{GSY} for a single black object, the analogous one for multi-centers will  turn out  a bit more involved again due to the relentless  cross-terms. Nevertheless with such a map in hand, transforming  amongst 4D/5D multi-center charges, we  verify that our fragmented harmonic functions are indeed direct 5D  lifts of  4D multi-center solutions.   This enables us to  confirm    commutativity of the  following box  diagram. 
\begin{eqnarray}
\xymatrix{ { \begin{tabular}{|c|}  \hline  4D \\ Black  Hole \\  \hline  \end{tabular} }  \ar[rrrr]^{Fragmentation}  \ar[ddd]_{4D/5D \; Map} &  &  &  & { \begin{tabular}{|c|}  \hline  4D  Multi-  \\ Black  Holes \\   \hline  \end{tabular} }  \ar[ddd]^{4D/5D \; Map}   \\  &  &  &  &  \\  &  &  &  &  \\  { \begin{tabular}{|c|}  \hline  5D \\ Black  Ring \\  \hline  \end{tabular} }  \ar[rrrr]_{Fragmentation}  &  &   &  &  { \begin{tabular}{|c|}  \hline  5D  Multi- \\ Black  Rings \\   \hline  \end{tabular} }    }
\label{1.3}
\end{eqnarray}
As had  already been hinted by in  \cite{DGOV} in context to  the 4D set-up;  eq.(\ref{1.3})  seems to predicate  the suggestion in 5D, that fragmentation might be thought of as a possible  recipe for generating classes of multi-center configurations once given corresponding single-center ones.  Of course the multi-rings that we generate in this note by these  methods, are by  no means any new solutions which  had previously been unheard of.  For that matter, we  point  to some of the extensive  literature, where several classes of  5D multi-center  solutions have  been worked out :   \cite{BW}, \cite{BW2}, \cite{BWW}, \cite{BGL}, \cite{C}.   The focus in this note is based more in the spirit of the box diagram in  eq.(\ref{1.3}) and studying the details therein.  

Whilst meandering amidst this impending scheme of things, we are duly  confronted with issues concerning the  physically meaningful  definition  of charges in 4D and 5D.  We begin with an apprehension of  the single black hole/ black ring duality  by  matching 4D two-center harmonics to 5D black ring harmonics.  Such a comparision invokes  symplectic charge transformations  going from 4D to 5D.  Additionally these 4D/5D transformations also make way for  an alternative  derivation of black ring angular momenta.  A  clear notion of single-center 4D/5D mapping, now equips us  to move  on  to study the interpretation of 5D multi-center charges. First we procure the  5D charge splitting equations via implementation of the 4D charge splitting equations as well as the single-center 4D/5D lift. The 5D equations so obtained definitely carry the baggage of cross-terms, due to the fact that  the 4D/5D transformations are non-linear in the dipole fields.   Moreover we shall see that it now becomes relevant to identify which of these charges  is of  Maxwell type and which of  Page type.  This discussion picks up from  \cite{HOT} and  continues  further for the case of fragmented charges.  In fact we shall see that in 5D the charges  ${{ Q_A}_i}_{(5D)}$  which actually engage in  fragmentation  are  Page charges. These are really the physical multi-ring charges and not the charges $\widetilde{Q}_{A_{ i \,(5D) } }$  in terms of which the     multi-black ring metric is usually expressed.   We also write down an explicit expression transforming between these two types of charges.  In due course  the multi-center 4D/5D dictionary falls in place.   

As an  application of  charge fragmentation methods described here,  we derive the total angular momentum of a system of non-concentric multi-black rings by simply starting from the angular momentum of a single black ring and making use of   5D charge splitting equations. As a check for our answer, we  reduce to the special case of concentric black rings in order to compare the our result with the well-known expression of Gauntlett and Gutowski  \cite{GG}, \cite{GG2}; and     yes, their result is correctly reproduced !

The alluring calls for a geometric interpretation of these fragmented rings underscore the  final act.  In a multi-ring background,   individual rings receive multiple spectral flow shifts due to  fluxes emanating from  split-charge centers;  thus coining the notion of 'split-spectral flows'.  Each ring may be thought of as sourcing  a Dirac string generated due to its magnetic flux. In a Taub-NUT base, these rings are stacked in order of increasing radius. Hence, say the $i^{th}$-ring; in addition to its own Dirac string; also encircles  Dirac strings sourced by each of the $( i - 1 )$ rings of smaller radius in the Taub-NUT base.  And going around Dirac strings is by no means a free ride. It costs large gauge transformations, which can have long-term consequences if Chern-Simons terms are involved as well. This is how spectral flows arise. Therefore the case of our  $i^{th}$-ring multi-timing that many Dirac strings will face a horde of spectral flow shifts to its initial brane charges. This will completely account for the physical split-charges of fragmented rings. Moreover, adding up all the split-spectral flows of all of our wandering fragmented rings correctly gives back the spectral flow of an unfragmented single black ring system, as it should.  This sheds light on a geometrical view of  the origin of multi-ring Page charges and their cross terms.  In fact such split-spectral flows divide the geometry into patches with locally defined gauge field potentials, such that adjacent patches are related upto gauge transformations.

The organisation of this note may not be the most exciting part of this text to read,  but still........   Section 2 provides a lay-out of the 4D multi-center black hole technology and comments on its physical interpretation as baby universes.  Section 3 handles  harmonics, charges and  angular momenta  of a single black ring in Taub-NUT  from a 4D/5D map.   Section 4 is where 5D fragmentation takes shape. We set-up  conditions for  black ring fragmentation  and provide an interpretation for multi-center 5D charges. This follows by writing down a multi-center 4D/5D charge dictionary and also deriving the angular momenta of (non-)concentric multi-black rings.  Section 5 seeks to unfold  a geometric perspective on the above via the notion of split-spectral flows.  Alas, we must wind up.....  that's why there's section 6, concluding  and  throwing  pointers at further     directions.

\section{A glance at 4D black hole fragmentation }
\label{sec2}
In this section we briefly sketch the set-up of 4D black hole fragmentation and its interpretation of baby universes following the approach of \cite{DGOV}.  The conceptual basis behind the idea of baby universes lies in the phenomenon of AdS fragmentation \cite{Br}, \cite{MMS}, which was proposed as an instanton process wherein a  single black hole, seen as an excitation in one vacuum  configuration,    tunnels to a multi-black hole state appearing as  an excitation in another vacuum.  The two vacua lie in the asymptotic limits of a ``Euclidean time'' co-ordinate, which is defined by an entropy functional $S(x)$.  From the Euclidean metric (obtained after a Wick rotation) the $AdS_2 \times S^2$ geometry is seen to flow to a   product geometry of  $\otimes_{i = 1}^n  AdS_2^i \times S^2_i$ (  to leading approximation  ).   

As an explicit representation  of multi-black hole configurations, the authors of  \cite{DGOV} make use of the well-known multi-center solutions of ${\cal N} = 2$ supergravity from \cite{BD}, \cite{DGR}, \cite{D}, \cite{D2}.  The  idea behind the  fragmentation procedure is that the black hole harmonic functions interpolate between the single-center harmonics; at asymptotic infinity $x \rightarrow \infty$; and the multi-center harmonics;   which are achieved  upon approaching the near-horizon limit. In fact, near the $i^{th}$-horizon when $x \rightarrow x_i$, the $i^{th}$-black hole dominates the solution. Therefore given a single-center solution and implementing the above  idea, one can set up an ansatz  for harmonic functions of   fragmented black holes. Additionally  charge conservation constrains  the  distribution of charges at  fragmented centers.  In \cite{DGOV} it was shown that the supergravity configuration  of \cite{BD}, \cite{DGR}, \cite{D}, \cite{D2} can indeed be realised in this way via AdS fragmentation of a single black hole.  For the sake of setting up notation as well as later reference, let us flash a quick glance at how this works. 

Consider the harmonic functions of a single black hole in 4D with magnetic charges $p^I$ and electric charges $q_I$ placed at  the spatial origin in   ${\mathbb R}^3$
\begin{eqnarray}
U^I (x) = \frac{p^I}{|x|} + u^I   \qquad  \qquad
V_I (x) = \frac{q_I}{|x|} + v_I 
\label{2.2}
\end{eqnarray}
here $I = 0, 1, .....$ denotes vector multiplet indices;  $x \in {\mathbb R}^3$;  and  $u^I$, $v_I$ are constants  determined at infinity.   In these co-ordinates the pole at $x  = 0$ is the location of the horizon which has the topology of a two-sphere $S^2$.  Another ingredient we will require is the entropy functional $S(x) \equiv  S \left[ U^I (x) , V_I (x) \right]$. This is a specific   polynomial function  of the harmonics and only at the horizon does it attain the value of the entropy. Elsewhere $S(x)$ freely flows between its asymptotic limits. This flow in $S(x)$ will induce the harmonic functions $U^I (x)$, $V_I (x)$ to  interpolate between single-center and multi-center solutions.  At asymptotic infinity  with $x \rightarrow \infty$,   a single black hole geometry with charges $p^I$, $q_I$ placed at the origin and harmonics given by eq.(\ref{2.2})  leads to  $S(x) \; \rightarrow  \; c$ ( a finite number ).  When these harmonics are inserted in the metric we get the well-known topology of $AdS_2 \times S^2$ and $S(x)$ enters this   near-horizon Bertotti-Robinson  metric  as the square of the  $AdS_2$ radius.  The idea of AdS fragmentation now proposes treating $S(x)$ as a Euclidean time direction. Then the $S(x) \; \rightarrow  \; \infty$ asymptote serves as another vacuum into which there exists a finite probability amplitude for a single black hole system to tunnel into a system of multi-black holes.  The most general solution for  harmonic functions,  which interpolate between these asymptotic vacua, looks like
\begin{eqnarray}
U^I (x) = \sum_{i = 1}^n \frac{p^I_i}{|x - x_i|} + u^I   \qquad  \qquad
V_I (x) = \sum_{i = 1}^n \frac{{q_I}_i}{|x - x_i|} + v_I 
\label{2.5}
\end{eqnarray}
where $U^I (x)$,  $V_I (x)$ now describe  a multi-black hole system with $n$ horizons located at centers $\{ x_i \}$. Charge splitting is subject to  the following constraints 
\begin{eqnarray}
\sum_{i = 1}^n  p^I_i = p^I  \qquad \mbox{and} \qquad \sum_{i = 1}^n {q_I}_i     =  q_I
\label{2.6}
\end{eqnarray}
To fully specify the solution additional integrability conditions are also required
\begin{eqnarray}
\left( p^I_i  V_I (x)  -   {q_I}_i  U^I (x) \right) \bigg|_{x  = x_i} = 0
\label{2.7}
\end{eqnarray}
which have to be evaluated at each horizon.  Now we can see how the above harmonic functions interpolate between single and multi-center geometries as follows :  at asymptotic infinity  $x  \rightarrow \infty$,  the harmonics in eq.(\ref{2.5}) reduce to  eq.(\ref{2.2}) ( by using the constraints in  eq.(\ref{2.6}) ) and $S(x) \; \rightarrow  \; c$; whereas at each $x  \rightarrow  x_i$, only the  $i^{th}$-summands  in  eq.(\ref{2.5}) dominate, describing multiple black holes located at $\{ x_i \}$ respectively and consequently giving  $S(x) \; \rightarrow  \; \infty$. Hence flowing  $S(x)$ from $c$ to $\infty$ describes an AdS geometry fragmenting into a multi-AdS geometry.  Eqs.(\ref{2.5}), (\ref{2.6}) and (\ref{2.7})  were originally derived as part of the multi-center ${\cal N} = 2$ supergravity solution of \cite{BD}. In \cite{DGOV} this solution has been interpreted as remnants of an AdS fragmentation process.

\section{Black ring from 4D/5D duality}
\label{sec3}
In this section we demonstrate how  charges, harmonics and angular momenta of a black ring can be determined purely in terms of a 4D/5D duality.  While the charges and harmonics are straightforward to get; an explicit expression for ring angular momenta obtained from  4D/5D lifting will serve to compliment the usual supergravity derivations discussed in the literature. 

We start by considering the following two-center system in 4D : 
\begin{eqnarray}
U^0 (x) \; \; = \; \;  \frac{1}{|x|} + u^0  \qquad  \qquad  \qquad
V_A (x) &=& \frac{q_{A (4D)} }{|x - x_0|} + v_A   \nonumber  \\
U^A (x) \; \; = \; \; \frac{p^A_{(4D)} }{|x - x_0|} + u^A    \qquad  \qquad
V_0 (x) &=& \frac{q_{0 (4D)}  }{|x - x_0|} + v_0  
\label{2.1}
\end{eqnarray}
which consists of a single D6 charge ( $p^0_{(4D)} = 1$  ) at the origin $x = 0$ ( in ${\mathbb R}^3$ );  and $p^A_{(4D)}$, $q_{A (4D)}$ and $q_{0 (4D)}$ respectively D4, D2, D0 charges, which form  a 4D black hole  at $x = x_0$. In  \cite{GSY2}, the 4D metric of this system is decompactified  to yield a 5D black ring in Taub-NUT. Instead of doing that, here we go for a more direct comparison; namely, showing that the 4D harmonics above will be identical to 5D black ring harmonics once they are expressed via 5D charges. For this we will require the  4D/5D charge transformations
\begin{eqnarray}
p^A_{(4D)} &=&  p^A_{(5D)}    \label{3.1}  \\
q_{A (4D)} &=&  \left(  Q_{A (5D)} - 3 D_{ABC} p^B_{(5D)} p^C_{(5D)} \right)
\label{3.2}
\end{eqnarray}
where $Q_{A (5D)}$ and $p^A_{(5D)}$  respectively will turn out to be black ring electric and magnetic charges.  We shall soon comment on their interpretation.   An additional ingredient required to specify eq.(\ref{2.1}) are the integrability conditions, which  yield        
\begin{eqnarray}
v_0 &=&  - \frac{q_{0 (4D)}}{L}   \label{3.3}   \\
q_{0 (4D)} &=& v_A  p^A_{(4D)} \left( \frac{1}{L}   +     \frac{4}{R^2_{TN}}\right)^{-1}
\label{3.4}
\end{eqnarray}
here $L$ denotes the radial distance $|x_0|$. The presence of a D6-brane leads to a geometric transition when lifting to M-theory, giving a Taub-NUT space in the uncompactified directions. Therefore $U^0 (x)$ becomes a harmonic function in Taub-NUT with  $u^0 = \frac{4}{R^2_{TN}}$ ( with $R_{TN}$ as the asymptotic radius of Taub-NUT  ).  $u^A$ remains arbitrary and can be set to zero. Putting all this together, the 4D harmonics above can indeed be compared to the known  Taub-NUT-black ring harmonics in the literature  \cite{BKW}\footnote{  Compared to \cite{BKW} we have scaled the $p^A_{5D}$ charge by a factor of $(-1)$.  }  ( see also   \cite{EEMR3}   )
\begin{eqnarray} 
H_{TN} (x) \; \; =  \; \; \frac{4}{R^2_{TN}} + \frac{1}{|x|}  \qquad  \qquad   
L_A (x) &=&  v_A + \frac{  {Q_A}_{(5D)} - 3 D_{ABC} {p^B}_{(5D)} {p^C}_{(5D)} }{|x - x_0|}   \nonumber \\
K^A (x)  \; \; =  \; \; \frac{ {p^A}_{(5D)} }{|x - x_0|} \qquad \qquad \quad  \; \; 
M (x) &=&  \frac{ J_{tube} }{L}  + \frac{ - J_{tube} }{|x - x_0|} 
\label{3.5}
\end{eqnarray}
where  $J_{tube} \equiv - q_{0 (4D)}$, which is determined from eq.(\ref{3.4}),  is indeed  the intrinsic ( not total ) angular momentum  of the ring along the $S^1$ circle and is the M-theory lift of the D0-charge.  Thus the harmonic functions of the  4D two-center system under consideration are exactly equivalent to those of a  5D black ring in Taub-NUT\footnote{ A black ring in ${\mathbb R}^4$  ( see \cite{EEMR}, \cite{EEMR2},  \cite{BK3},  \cite{B}   )  can be extracted as a special case of eq.(\ref{3.5})  by  taking the limit  $R_{TN} \rightarrow \infty$. The conventions of \cite{EEMR}, \cite{EEMR2}  differ from \cite{BKW} by rescaling of charges; in this note we continue using the latter.  }.

Note that  these functions in  eq.(\ref{3.5})  ( along with integrability conditions  )  completely specify the black ring solution.  For the sake of completeness, let us quickly demonstrate   how this comes about.   Consider  the most general 5D ${\cal N} = 1$ ungauged supergravity solution  \cite{GGHPR},  \cite{GG3}    which is given  by the  following 5D metric and  gauge fields
\begin{eqnarray} 
ds^2_{5} &=& -  \; f^2  \; (  \; dt \; + \;  \omega  \; )^2 \;  + \;  f^{-1} \; ds^2 ( M_4 )  \nonumber \\
F^A &=&  d \left[ \; f  \; X^A \; ( \; dt  \; +  \; \omega \; )  \;  \right]  \; -  \;  \frac{2}{3}  \; f \; X^A  \;   (  \; d \omega   \; +   \; \star d \omega  \; ) 
\label{3.5a1}
\end{eqnarray}
where \;  $X^A$ \; are scalar fields in abelian vector  multiplets; they satisfy the constraint equation  $D_{ABC} X^A X^B X^C =1$ and $X_A$ are defined by the condition  $X^A X_A = 1$.   $ds^2 ( M_4 )$  in the  equation above refers to the Gibbons-Hawking metric of a 4D hyper-Kahler base space, which in our case is simply taken to be  $ds^2 ( TN )$, the Taub-NUT metric ( or  $ds^2 ( {\mathbb R}^4 )$  when considering  a black ring in flat space  ). Let $r$, $\theta$, $\phi$,  $\psi$ denote coordinates on the 4D base space with $( r, \theta, \phi )$ locally parameterising  an  ${\mathbb R}^3$ and $\psi$ running  along a compact $S^1$ with periodicity $4 \pi$.  The Hodge dual $\star$ is taken with respect to the 4D base space.   The function $f$ and one-form $\omega$ are fully nailed down in terms of four yet-to-be-specified  harmonic functions as follows
\begin{eqnarray} 
f^{-1}  \; X_A  &=&  \frac{1}{4}  \;  {H_{TN}}^{-1} \; D_{ABC} K^B K^C   \; +  \;  L_A   \nonumber \\
\omega &=& (  \; -  \frac{1}{8} \;  {H_{TN}}^{-2}  \;  D_{ABC} K^A K^B K^C   \;  -  \;  \frac{3}{4} {H_{TN}}^{-1}  \; K^A  \;  L_A  \; +  \;  M   \; )   \nonumber \\   &\times&   (  \; d \psi  \;  +  \;  cos \; \theta  \; d \phi  \;  )  \;  +  \;  \widehat{\omega}
\label{3.5a2}
\end{eqnarray}
The notation used in this equation is intentionally suggestive. Furthermore, $\widehat{\omega}$ is defined by
\begin{eqnarray} 
\nabla \times \widehat{\omega}   \; =  \;  H_{TN}  \; \nabla M   \; -  \;  M  \;  \nabla H_{TN}  \; +  \; \frac{3}{4}  \left(  L_A  \; \nabla K^A   \;  -  \;  K^A   \; \nabla  L_A   \right)
\label{3.5a3}
\end{eqnarray}
Now inserting the explicit form of the harmonic functions of  eq.(\ref{3.5}) into eqs.(\ref{3.5a1}),  (\ref{3.5a2}) and (\ref{3.5a3})  simply  reproduces the  complete  black ring solution of \cite{BKW} in  Taub-NUT  ( or \cite{EEMR3} in  ${\mathbb R}^4$  ).  Moreover,  operating  the gradient on both sides of  eq.(\ref{3.5a3}) and evaluating at the poles,  exactly  recovers  the   integrability conditions of eq.(\ref{2.7}), which are subsequently solved to get eqs.(\ref{3.3}) and (\ref{3.4}).  This prescription goes through for multi-rings as well. Inserting  appropriate multi-ring harmonics into the same 5D supergravity metric given above, one can recover the multi-black ring solution \cite{GG}, \cite{GG2}. In this sense, the harmonics and integrability conditions can be said to be sufficiently representative of the solutions  of single as well as multi-black rings.  For what follows here, we shall adopt this stance as well. Therefore the focus in this note shall not be on solving supergravity equations themselves, but rather on obtaining quantities such as multi-ring harmonics, charges and angular momenta from  ring fragmentation and spectral flows.

Now coming back to the 4D/5D transformations, a  comment on  eqs.(\ref{3.1}) and (\ref{3.2}) is due. These equations were derived in \cite{GSY}  by considering symplectic shifts in electric charges due the presence of a magnetic flux such that the  degeneracy of microstates  remains invariant. Subsequently this leads to matching of leading order entropies for 4D and 5D black holes. Also, the authors of  \cite{BKW}, \cite{BK2},  \cite{BK4}  further clarify these transformations when  interpolating  from a  4D black hole to a  5D black ring. While $q_{A (4D)}$ is the observable in 4D, from the 5D perspective it is $Q_{A (5D)}$ which is the observed charge. Let  us point out to yet another interpretation of these transformations coming  from spectral flow shifts ( as in \cite{BCDMV} ) associated to the 5D Chern-Simons term. In a later section, we pursue this last observation further. 

Much like the above-mentioned D2 charges, there also occurs a shift for  D0 charges ( again due to \cite{GSY}  )
\begin{eqnarray} 
q_{0 (5D)}  =  q_{0 (4D)} - ( p^A_{(4D)} q_{A (4D)}  + D_{ABC} p^A_{(4D)}  p^B_{(4D)} p^C_{(4D)}  )
\label{3.7}
\end{eqnarray}
Starting from this relation we now obtain an independent identification of the total black ring angular momenta. Simply using eqs.(\ref{3.1}), (\ref{3.2}), (\ref{3.3}) and (\ref{3.4}) into eq.(\ref{3.7}) yields 
\begin{eqnarray} 
q_{0 (5D)}  =  v_A  p^A_{(5D)} \left( \frac{1}{L} + \frac{4}{R^2_{TN}}\right)^{-1}     \hspace{-0.3cm}      - p^A_{(5D)} ( Q_{A (5D)} - 2 D_{ABC} p^B_{(5D)} p^C_{(5D)} )
\label{3.10}
\end{eqnarray}
Now let us  denote  $q_{0 (5D)} \equiv  - \frac{G}{3 \pi} J_{\psi}$, where $G$ is the 5D Newton's constant.  Then   $J_{\psi}$    exactly compares to the total angular momentum of the ring along the $S^1$ circle as given in  \cite{BKW} ( or \cite{EEMR}  on reducing to ${\mathbb R}^4$  ).  The first term of $J_{\psi}$ is the  intrinsic  angular momentum arising  via the presence of D0 charges along the $S^1$ circle ( $\psi$-direction  ); the second  component  describes the angular momentum  induced in the presence of a magnetic flux. In addition there is yet another angular momentum characterising the ring;  one  associated to the $\phi$-circle  along the $S^2$,  perpendicular to the  $\psi$-circle. In the absence of  D0 charges along the $\phi$-circle, with  only  flux going through it, the  angular momentum  contribution ( denote as $J_{\phi}$ ) is solely flux-induced, thus giving    
\begin{eqnarray} 
J_{\phi} =  J_{\psi} -  \frac{3 \pi}{G} J_{tube}
\label{3.11.2}
\end{eqnarray} 
Thus far we  conclude that  explicit application of the 4D/5D correspondence   correctly identifies the charge prescription, harmonic functions as well as angular momenta of a black ring. Proceeding this way the leading order black ring entropy too can be obtained, as well as its one-loop correction. Since the references  \cite{BK2},  \cite{BK4}, \cite{GSY2}, \cite{CGMS} do justice to the former and  \cite{BK} to the latter, we shall have no more to say on that. Equipped with these tools, we shall next test their application for the case of multi-center black holes/rings.

\section{Black ring fragmentation and 5D charge splitting}
\label{sec4}
As seen in section 2,  4D charge fragmentation is given by simple linear relations in terms of fragmented charges. For D4, D2, D0 branes respectively, we denote these splittings as  
\begin{eqnarray}
\sum_{i = 1}^n  {p^A_i}_{(4D)} = {p^A}_{(4D)}  \qquad  \sum_{i = 1}^n {{q_A}_i}_{(4D)} = {q_A}_{(4D)}   \qquad  \sum_{i = 1}^n {{q_0}_i}_{(4D)} = {q_0}_{(4D)}
\label{4.1}
\end{eqnarray}
Let us note that in 4D these are also the physically observed charges. We would now like to construct the analog of these equations in 5D.  In that case, as we shall soon see,  the charge fragmentation equations  are not only non-linear (in the dipole charges) but also involve cross-term contributions arising from multiple charge centers. 

\subsection{5D multi-black ring charges and harmonic functions from  fragmentation}
Owing to the trivial 4D/5D relation for magnetic charges ${p^A}_{(5D)}$ ( as in eq.(\ref{3.1}) ), their  splitting into 5D components is straightforward.
\begin{eqnarray}
{p^A}_{(5D)} = \sum_{i = 1}^n  {p^A_i}_{(5D)}
\label{4.3}
\end{eqnarray}
The more interesting case is that of  the electric charge ${Q_A}_{(5D)}$ of a  single black ring.  Since  this charge differs from the corresponding 4D  charge $q_{A (4D)}$ by large gauge transformations induced via the  Chern-Simons term in the 5D action, therefore the 5D splitting for ${Q_A}_{(5D)}$ will turn out to be more involved.  Analogous to the 4D case, let us define the fragmentation
  of this charge  to be
\begin{eqnarray}
  {Q_A}_{(5D)}  \, \equiv \,  \sum_{i = 1}^n  {{Q_A}_i}_{(5D)}
\label{4.5}
\end{eqnarray}
where we now have to determine  ${{Q_A}_i}_{(5D)}$ and then provide it with  a physical interpretation.  To do this, we substitute the conditions given in  eq.(\ref{4.1}) into  eqs.(\ref{3.1}) and  (\ref{3.2}).  Upon further rearranging we get  
\begin{eqnarray}
\hspace{-1cm}  {Q_A}_{(5D)}  &=&    \left\{ \sum_{i = 1}^n {{q_A}_i}_{(4D)}  +  \sum_{i = 1}^n  \sum_{j = 1}^n  3 D_{ABC}  {p^B_i}_{(4D)} {p^C_j}_{(4D)} \right\}     \label{4.2a}  \\    &=&    \sum_{i = 1}^n  \left\{ \left(  \widetilde{Q}_{A_{ i \,(5D) } }    - 3 D_{ABC}  {p^B_i}_{(5D)} {p^C_i}_{(5D)}  \right)   +    \sum_{j = 1}^{n}  3  D_{ABC}    {p^B_i}_{(5D)} {p^C_j}_{(5D)}  \right\}    
\label{4.2}
\end{eqnarray}
where the last line has  been converted to  5D quantities with the intent of extracting 5D charge fragments.      $\widetilde{Q}_{A_{ i \,(5D) } }$  is introduced as a  new 5D variable defined by the following 4D/5D transformation
\begin{eqnarray}
\widetilde{Q}_{A_{ i \,(5D) } }  =  {{q_A}_i}_{(4D)}  +  3 D_{ABC}  {p^B_i}_{(4D)}  {p^C_i}_{(4D)}
\label{4.12}
\end{eqnarray}
Notice that the right-hand side of eq.(\ref{4.2}) has been expressed in a way that facilitates comparison to the literature.   $\widetilde{Q}_{A_{ i \,(5D) } }$ is actually a 5D charge associated to the $i^{th}$ black ring and  is the one that appears  in  the usual  5D multi-ring supergravity solutions  \; ( for instance see \cite{GG}, \cite{GG2}  ). \;  In this way, eq.(\ref{4.2}) is simply  the ADM mass\footnote{ Even though  \cite{GG}, \cite{GG2} only refer to concentric rings, the above comparison is still meaningful because effects due to  non-concentricity only start showing up for quantities involving the position vector $\vec{L}$, such as angular momentum, entropy, etc.   }   of  \cite{GG}, \cite{GG2}.  Note that  because  these references dwell  in conventions different from ours, the following rescaling of charges must be used : ${p^A_i}_{(5D)}  \longrightarrow  \sqrt{2} {p^A_i}_{(5D)}$. Also they use $C_{ABC}$ as the intersection number, which relates to the one used  here via  $C_{ABC} = 6  D_{ABC}$.  

Despite the above comparison, let us remark  that  in our case   eq.(\ref{4.2}) is obtained as a result of 5D fragmentation. Therefore  it is clear that  summing all the $\widetilde{Q}_{A_{ i \,(5D) } }$'s  over all $i$ would not  conserve  ${Q_A}_{(5D)}$.  The charges that are actually   involved in 5D fragmentation are clearly the ${{Q_A}_i}_{(5D)}$'s  and not $\widetilde{Q}_{A_{ i \,(5D) } }$'s.  So the question arises, which of these two is the correct physical observable ?  In order to answer this, we shall take a closer look at the interpretation of each of these charges via their integral representations. It will turn out that it is in fact the ${{Q_A}_i}_{(5D)}$'s that are   the physically observable quantities  and not  the  $\widetilde{Q}_{A_{ i \,(5D) } }$'s.  The subtlety between ${{Q_A}_i}_{(5D)}$ and $\widetilde{Q}_{A_{ i \,(5D) } }$ arises precisely  due to the presence of cross-terms relating different charge centers.  The consequences of these cross-terms  will also be evident  in other  quantities such as   multi-ring angular momenta.  For later reference, let us note down the relation between the two  charges
\begin{eqnarray}
{{Q_A}_i}_{(5D)}  =  \widetilde{Q}_{A_{ i \,(5D) } }  + \sum_{  j = 1  }^{ i - 1 }  3 D_{ABC} \left(  {p^B_i}_{(5D)} {p^C_j}_{(5D)} +  {p^B_j}_{(5D)} {p^C_i}_{(5D)}   \right)  
\label{4.4}
\end{eqnarray}
Whilst plucking  this expression   from  eq.(\ref{4.2})  one has also to keep in mind that ${{Q_A}_i}_{(5D)}$ should be independent of how the cycles $B$ and $C$ have been labelled. Therefore the resulting  expression for ${{Q_A}_i}_{(5D)}$  has to be symmetrised as done above.

Now let us try to  understand the various 5D charges discussed above in the form of  integrals over near-horizon patches.   In \cite{HOT} it was shown that in terms of purely near-horizon fields (and not requiring data from the complete solution) of a single black ring, the charge  ${Q_A}_{(5D)}$ can be understood as a Page charge rather than  a  Maxwell charge (  see also \cite{DM} for a clear exposition on the different notions of charge  )
\begin{eqnarray}
{Q_A}_{(5D)} =  \int_{\Sigma} \left( \star  a_{AB} F^B + 3 D_{ABC} A^B \wedge F^C  \right)
\label{4.7}
\end{eqnarray}
where the range of integration, denoted by $\Sigma$, is  a spatial 3-cycle in the vicinity of the black ring horizon. The  $a_{AB}$, which is a function of the scalar moduli, serve the usual purpose of lowering vector multiplet indicies.    $A^B$ denote  near-horizon $U(1)$ gauge fields around the black ring.  A Page charge is conserved,  localised and quantised, but not gauge invariant.  The near-horizon integral on the right-hand side of eq.(\ref{4.7})  implicitly represents a Page charge.  In \cite{HOT}, they explicitly compute  this  integral and  show that it indeed results in  the black ring charge ${Q_A}_{(5D)}$.
 
Adapting the results of \cite{HOT} to the present context of fragmented rings, we now argue that the ${{Q_A}_i}_{(5D)}$'s  are also Page charges. This is consistent with the role of eq.(\ref{4.5})  as  a charge conservation equation.  Then   ${{Q_A}_i}_{(5D)}$ should also have an  expression as a localised charge resulting from a  near-horizon integral
\begin{eqnarray}
{{Q_A}_i}_{(5D)}  \; \; \mathop{=}\limits^{?} \; \;  \int_{\Sigma_i} \left( \star  a_{AB} F^B_i + 3 D_{ABC} A^B_i \wedge F^C_i  \right)
\label{4.8}
\end{eqnarray}
for $A^B_i$ as $U(1)$ gauge fields locally defined  in the neighbourhood of the $i^{th}$-ring horizon.  $\Sigma_i$  denotes a  3-cycle enclosing the $i^{th}$-horizon and $F^B_i = d A^B_i$.   So the question then is : does this  integral in   eq.(\ref{4.8}) work out to give   ${{Q_A}_i}_{(5D)}$ ?  Upon inserting the following expression for the gauge field :
\begin{eqnarray}
A^B_i =  &-&  \left[  \left(  D^{BC}  \left(  \widetilde{Q}_{A_{ i \,(5D) } }  - 3  D_{CDE} {p^D_i}_{(5D)} {p^E_i}_{(5D)}  \right)  + 2 \sum_{j = 1}^{n}  {p^B_j}_{(5D)}    \right) d \psi \right.  \nonumber \\   &\,&  \quad  +  \left.  \left( {p^B_i}_{(5D)} (1 + x) + 2 \sum_{j = i + 1}^{n}  {p^B_j}_{(5D)}  \right) d \chi  \right]
\label{4.9}
\end{eqnarray}
into the  integral in eq.(\ref{4.8}), the authors of \cite{HOT} indeed do obtain   the  expression\footnote{ In \cite{HOT}  the  computation was  first done for the special case  of only one  vector field, and then it was generalised to $n$ $U(1)$ fields by simply carrying through the same calculation with vector indicies.}   we had in  eq.(\ref{4.4}).   In eq.(\ref{4.9}), the variables $\psi$, $\chi$ and $x$ are the usual ring-coordinates  ( notation follows from \cite{EEMR} ). $( \psi + \chi/2 )$ and $x$ parametrise  the $S^2$, while $( \psi - \chi/2 )$ runs along the $S^1$   near the horizon  of the $i^{th}$-black ring.    The gauge fields $A^B_i$ are locally defined patchwise. Gluing of adjacent patches is achieved via gauge transformations. In eq.(\ref{4.9}), $i = 1$ refers to the innermost ring ( smallest radius ) and the radial parameter monotonically increases with increasing $i$. The expression for $A^B_i$ used in \cite{HOT} was extracted from the supergravity  solution  of \cite{GG},  \cite{GG2} for concentric black rings.  The same is reliable for non-concentric rings too, since restrictions to concentricity mainly become relevant when evaluating integrability conditions ( and those bear consequences for multi-ring angular momenta ).

From the expression for  $A^B_i$ above, we see that the gauge field around the $i^{th}$-ring  also feels the back-reaction due to  dipole fields  from neighbouring rings. It is precisely this dipole field back-reaction that  leads to  cross-terms  in the computation of ${{Q_A}_i}_{(5D)}$.  In our case we tried to derive  these terms  from the construction of 5D fragmentation. It is gratifying to note that they exactly compare with those coming from the integral   of   \cite{HOT}.  As we shall see that fragmentation of a single ring indeed does reproduce  the correct  multi-ring charges.

Now turning our attention to  $\widetilde{Q}_{A_{ i \,(5D) } }$, let us see why this is in fact not a physical charge.  From the definition of  $\widetilde{Q}_{A_{ i \,(5D) } }$ in eq.(\ref{4.12}), its  4D/5D transformation  is identical to that of a single black ring system  with electric charge $\widetilde{Q}_{A_{ i \,(5D) } }$ and magnetic charge ${p^A_i}_{(5D)}$. This is in stark contrast to the analogous  transformation of  ${{Q_A}_i}_{(5D)}$ (  which  can be read-off from  eq.(\ref{4.2a})  ). Unlike   ${{Q_A}_i}_{(5D)}$, we see that $\widetilde{Q}_{A_{ i \,(5D) } }$ clearly does not sense the background  reaction due to  neighbouring rings. Hence such a charge cannot be given a global physical meaning in a multi-ring geometry. Its presence is at best only a local approximation. Therefore its integral representation is trivially identical to eq.(\ref{4.7}) after all charges ( which enter into the explicit expressions for the gauge potentials  )  have been replaced by those at the $i^{th}$-center.

Eqs.(\ref{4.2a}) and (\ref{4.12})   essentially describe  the multi-center 4D/5D dictionary  for electric  charges.  As expected the physical multi-center Page charge ${{Q_A}_i}_{(5D)}$  transforms in a more complicated way than  ${Q_A}_{(5D)}$ (eq.(\ref{3.2})), due to  the multi-black ring background. On  the other hand, the charges $\widetilde{Q}_{A_{ i \,(5D) } }$, though unphysical,  retain manifest symplectic invariance of the original single-center solution. Each of the  $\left( {p^A_i}_{(5D)} \, , \, \widetilde{Q}_{A_{ i \,(5D) } }   \right)$    manifestly transform  as a symplectic pair. This underlying property often makes it  convenient  to express multi-black ring solutions in terms of these charges  \,  ( \, as has   been usual practice in the literature \, ).

Having explicitly constructed the 5D charge fragmentation equations for  magnetic and  electric  charges \; $\left( \, {p^A}_{(5D)} \, , \, {{Q_A}}_{(5D)} \, \right)$ \;  along with the relevant multi-center 4D/5D transformations, we  are now equipped to derive two of the multi-ring harmonic functions  $\left(  K^A (x) , L_A (x)  \right)_{multi}$  from the single-ring harmonics  $\left(  K^A (x) , L_A (x)  \right)_{single}$  by  merely implementing  the fragmentation recipe of section 2. 
As in eqs.(\ref{2.5}) and  (\ref{2.6})  we have 
\begin{eqnarray}
L_A (x) \bigg|_{single} &=&  v_A + \frac{  {Q_A}_{(5D)} - 3 D_{ABC} {p^B}_{(5D)} {p^C}_{(5D)} }{|x - x_0|}   \nonumber   \\ 
&\longrightarrow&  v_A + \sum_{i = 1}^n \frac{ {{Q^{\bullet}_A}_i}_{(5D)} - 3 D_{ABC}  {{p^{\bullet}}^B_i}_{(5D)}  {{p^{\bullet}}^C_i}_{(5D)}  }{|x - x_i|} =  L_A (x) \bigg|_{multi}
\label{4.17}
\end{eqnarray}
which is subject to the constraint 
\begin{eqnarray}
{Q_A}_{(5D)} - 3 D_{ABC} {p^B}_{(5D)} {p^C}_{(5D)} =  \sum_{i = 1}^n  \left(  {{Q^{\bullet}_A}_i}_{(5D)}  -  3 D_{ABC}  {{p^{\bullet}}^B_i}_{(5D)}  {{p^{\bullet}}^C_i}_{(5D)}  \right)
\label{4.18}
\end{eqnarray}
Eqs.(\ref{4.17})   and (\ref{4.18})  constitute a natural 5D fragmentation ansatz   with newly-defined charges  ${{Q^{\bullet}_A}_i}_{(5D)}$ and  ${{p^{\bullet}}^A_i}_{(5D)}$  such that at $x \rightarrow \infty$    one recovers  $L_A (x) \bigg|_{single}$   while  at $x \rightarrow  x_i$  the solution (at leading approximation) appears like a single black ring at the $i^{th}$ location. Now  the constraint in eq.(\ref{4.18}) above  is identical in form to the charge splitting eq.(\ref{4.2}), which  suggests the identification
\begin{eqnarray}
{{Q^{\bullet}_A}_i}_{(5D)} \equiv \widetilde{Q}_{A_{ i \,(5D) } }  \qquad   {{p^{\bullet}}^A_i}_{(5D)}  \equiv  {p^A_i}_{(5D)} 
\label{4.19}
\end{eqnarray}   
From this we also see  how the charges $\widetilde{Q}_{A_{ i \,(5D) } }$   enter into  the 5D harmonics and subsequently into the metric. Of course the above harmonic function could also have been written in terms of   ${{Q_A}_i}_{(5D)}$, but then  the expressions would only get a little messy as we proceed.

Another remark that we can make at this stage is that eq.(\ref{4.17})  (  along with the conditions in eqs.(\ref{4.18}) and (\ref{4.19})  )  could also have been  obtained  via a different route; namely,  by direct use of  the multi-ring 4D/5D transformation  (  eq.(\ref{4.12}) )  into eqs.(\ref{2.5}) and (\ref{2.6}).  This is consistent with the commutativity  of the diagram in eq.(\ref{1.3}), which suggests that fragmenting a single black ring into multiple black rings reproduces the same configuration  as that obtained by a direct 5D lift of the appropriate  4D multi-center black holes.

Dealing with  the harmonic function $K^A (x)$  for magnetic charges is now straightforward :
\begin{eqnarray}
K^A (x) \bigg|_{single} &=& \frac{ {p^A}_{(5D)} }{|x - x_0|}   \nonumber   \\     &\longrightarrow&   \sum_{i = 1}^n \frac{ {p^A_i}_{(5D)}  }{|x - x_i|} =  K^A  (x) \bigg|_{multi}
\label{4.20}
\end{eqnarray}   
which is again subject to 
\begin{eqnarray}
{p^A}_{(5D)} =  \sum_{i = 1}^n  {p^A_i}_{(5D)}
\label{4.21}
\end{eqnarray}  

As per the other two black ring harmonic functions $H_{TN} (x)$ and $M (x)$ : the former remains unchanged under fragmentation as our brane configuration includes only a single D6 charge ( which lifts to a Kaluza-Klein monopole in 5D  );  while fragmentation of the latter comes up in the following sub-section.

\subsection{Multi-black ring  angular momenta  from  ring  fragmentation}
We are now ready to derive the expressions for angular momenta of a multi-black ring system from 5D fragmentation techniques. Our starting point is eq.(\ref{3.10}) : the angular momentum of a single black ring along the $\psi$-direction 
\begin{eqnarray}
J_{\psi} =  \frac{3 \pi}{G}  J_{tube} +  \frac{3 \pi}{G} p^A_{(5D)} (  Q_{A (5D)} -  2 D_{ABC} p^B_{(5D)} p^C_{(5D)} ) 
\label{4.22}
\end{eqnarray}   
Inserting the 5D charge splitting eqs.(\ref{4.2}) and (\ref{4.3})  into the above we readily  obtain 
\begin{eqnarray}
J_{\psi} \, =  \, \frac{3 \pi}{G}  \sum_{i = 1}^n  J_{tube}^i  &+&    \frac{3 \pi}{G}  \left[  \sum_{i, j = 1}^n  {{p}^A_i}_{(5D)}   \left(  \widetilde{Q}_{A_{ i \,(5D) } }  -  3 D_{ABC}  {{p}^B_j}_{(5D)}  {{p}^C_j}_{(5D)}  \right)  \right.  \nonumber \\     &\,&  \qquad  +   \left.   \sum_{i, j, k  = 1}^n   D_{ABC}  {{p}^A_i}_{(5D)}   {{p}^B_j}_{(5D)}  {{p}^C_k}_{(5D)}  \right]
\label{4.23}
\end{eqnarray}  
where the quantities $J_{tube}^i$  have yet to be  determined from integrability  conditions. As  a special case of our result in eq.(\ref{4.23}),  we shall  be able to  reproduce   the  expression for  angular momentum  of  concentric black rings which was first derived by  Gauntlett and Gutowski  in  \cite{GG}, \cite{GG2}  in the context of 5D supergravity.   In order to obtain  $J_{tube}^i$,   we will first have to determine  the multi-ring  harmonic function $M (x)$, from where  $J_{tube}^i$ can be extracted.   Therefore,  fragmenting  the function  $M (x)$ yields 
\begin{eqnarray}
M (x) \bigg|_{single} &=&  v_0  +  \frac{  - J_{tube}  }{|x - x_0|}   \nonumber   \\     &\longrightarrow&   v_0  +  \sum_{i = 1}^n \frac{ - J_{tube}^i  }{|x - x_i|} =  M  (x) \bigg|_{multi}
\label{4.24}
\end{eqnarray}  
subject to 
\begin{eqnarray}
J_{tube} =  \sum_{i = 1}^n  J_{tube}^i
\label{4.25}
\end{eqnarray}  
Additionally,   the multi-ring harmonics $(  H_{TN}(x), K^A(x), L_A(x), M(x)  )_{multi}$   above   also have  to satisfy  integrability conditions as in eq.(\ref{2.7}).  These are to be evaluated  at each horizon.  Starting with    $x =  0$,  we get 
\begin{eqnarray}
v_0  =    \sum_{i = 1}^n  \frac{ J_{tube}^i  }{L_i} 
\label{4.26}
\end{eqnarray}  
This determines the constant $v_0$ in terms of $J_{tube}^i$ ( which we still have to fix in terms of electric and magnetic charges  ) and $L_i$ ( which is the radial distance in ${\mathbb R}^3$ of the $i^{th}$ pole from the origin  ). However, as discussed earlier, $v_0$ is a constant predetermined  at infinity and should not be affected by the process of fragmentation. As a  consistency check we shall see in what follows  that  eq.(\ref{4.26}) is indeed  identical to eq.(\ref{3.3}) obtained earlier in section 3.  Before that we will require to  compute the remaining $n$  conditions at the  horizons  $\{ x_i  \}$. This yields 
\begin{eqnarray}
- J_{tube}^i  &=&  \left( \frac{4}{R^2_{TN}} + \frac{1}{L_i}  \right)^{-1}   \left( {p^A_i}_{(5D)} v_A   +   \sum_{ \begin{array}{c} {j = 1} \\ j \neq i \end{array} }^n  \frac{{p^A_i}_{(5D)}  \left(  \widetilde{Q}_{A_{ i \,(5D) } }  -  3 D_{ABC}  {{p}^B_j}_{(5D)}  {{p}^C_j}_{(5D)}  \right)     }{ \sqrt{L_i^2 - 2 L_i L_j cos \theta_{ij} + L_j^2} }    \right. \nonumber \\   &-& \left.  \sum_{ \begin{array}{c} {j = 1} \\ j \neq i \end{array} }^n  \frac{  \left(  \widetilde{Q}_{A_{ i \,(5D) } }  -  3 D_{ABC}  {{p}^B_i}_{(5D)}  {{p}^C_i}_{(5D)}  \right)   {p^A_j}_{(5D)}  }{ \sqrt{L_i^2 - 2 L_i L_j cos \theta_{ij} + L_j^2} }  \right)
\label{4.27}
\end{eqnarray}  
where $\theta_{ij}$ is the angle between $\vec{L_i}$, $\vec{L_j}$ $\in$ ${\mathbb R}^3$. Now rearranging eq.(\ref{4.27}) for $\frac{ J_{tube}^i  }{L_i}$ and then  inserting back into  eq.(\ref{4.26})  produces 
\begin{eqnarray}
v_0  =  - \frac{4 \, J_{tube}}{R^2_{TN}} -   v_A  {p^A}_{(5D)}
\label{4.28}
\end{eqnarray}  
after also using eqs.(\ref{4.3}) and (\ref{4.25}).   Indeed eq.(\ref{4.28})  is  precisely  the value of $v_0$   obtained earlier by inserting eq.(\ref{3.4}) into eq.(\ref{3.3}).  

Now with eqs.(\ref{4.27}) and (\ref{4.28})  the function $M  (x) \bigg|_{multi}$  is fully specified.  Thus simply from 5D black ring fragmentation we were able to construct all of the multi-black ring harmonic functions.  Moreover inserting   eq.(\ref{4.27})  for $J_{tube}^i$  into eq.(\ref{4.23})  results in  the complete expression for the  total multi-black ring angular momentum  in the $\psi$-direction :  $J_{\psi}$.  Also the angular momentum in the $\phi$-direction :  $J_{\phi}$, can  then  be  read-off from  $J_{\psi}$  since 
\begin{eqnarray}
J_{\phi} = J_{\psi} -  \frac{3 \pi}{G}  \sum_{i = 1}^n  J_{tube}^i
\label{4.29}
\end{eqnarray}  
still continues to hold. 

An additional comment on eq.(\ref{4.27}) is due.  Let us take a closer look at   the last two terms on the right-hand side of this equation.  As long as the multi-center charges are constrained to remain  mutually non-local, then   $\vec{L_i} \neq \vec{L_j}$ will hold and that avoids any potential singularity in eq.(\ref{4.27}). Hence the sum of the two  numerators ( within the summation symbols  )  is allowed to assume  any non-zero value. From the 4D  point of view,  this is precisely the condition for the dual 4D     charges  $( {p^A_i}_{(4D)}, \, {{q_A}_i}_{(4D)} )$   to be non-parallel ( on the charge lattice  ). This was the  interesting  new feature in the multi-center solution  of \cite{BD}, \cite{DGR}, \cite{D}, \cite{D2}.  On the other hand,  if the condition $\vec{L_i} \neq \vec{L_j}$ were to be relaxed; then we  would be  required to impose 
\begin{eqnarray}
{p^A_i}_{(5D)}  \left(  \widetilde{Q}_{A_{ i \,(5D) } }  -  3 D_{ABC}  {{p}^B_j}_{(5D)}  {{p}^C_j}_{(5D)}  \right)  -  \left(  \widetilde{Q}_{A_{ i \,(5D) } }   -  3 D_{ABC}  {{p}^B_i}_{(5D)}  {{p}^C_i}_{(5D)}  \right)   {p^A_j}_{(5D)} = 0   \hspace{0.7cm}
\label{4.30}
\end{eqnarray}
for all $i \neq j$, thereby eliminating the last two terms in eq.(\ref{4.27}).   The  corresponding  4D charge vectors  $( {p^A_i}_{(4D)}, \, {{q_A}_i}_{(4D)} )$  are now parallel-aligned  on the charge lattice\footnote{Note that being parallel on the charge lattice should not be confused with co-linearity of the poles in ${\mathbb R}^3$.  Even for parallel charges  the multi-center  poles  are still  free to remain non-colinear. From a  4D/5D perspective, non-colinear  D4-D2-D0 poles in 4D lift  to non-concentric rings in 5D.   }.   The reason we made the above comment is because the construction in  \cite{GG}, \cite{GG2}  does restrict to  eq.(\ref{4.30}) and hence we too will need to make use of it whenever comparing to their results. For all other purposes, our results continue to hold for   non-parallel charges in general.     

Eq.(\ref{4.23}) along with eq.(\ref{4.27}) gave us the most general result for the angular momentum ( along the $\psi$-coordinate  ) of  non-concentric multi-black rings. We would now like to reduce our  result to  the case of concentric rings so as to compare it with the well-known answer of  \cite{GG}, \cite{GG2}, which was derived  using 5D supergravity techniques of  \cite{GGHPR} and  \cite{GG3}.  First we set all angles $\theta_{ij}$ between the poles to zero.  The  co-linear  alignment of poles in  ${\mathbb R}^3$ translates to concentric  rings in 5D. In order to eliminate Dirac-Misner strings,  \cite{GG}, \cite{GG2} choose to impose eq.(\ref{4.30}), which can be interpreted as a restriction to parallel 4D charge vectors\footnote{  In fact this is not the most general way to eliminate Dirac-Misner strings and admittedly  ends up making the solution of  \cite{GG}, \cite{GG2}  highly restrictive. In general it suffices to impose the integrability conditions as we have done in this note. The difference with  \cite{GG}, \cite{GG2} is that those authors impose eq.(\ref{4.27}) in a very special way.  }.   From our discussion above, we see that it is still possible to continue with non-parallel charges by trading-off mutual locality of charges.   Nevertheless to make contact  with  \cite{GG}, \cite{GG2}; we use eq.(\ref{4.30})  in eq.(\ref{4.27}) with all angles $\theta_{ij} = 0$ and thus arrive at the desired result upon plugging everything back into eq.(\ref{4.23}). To facilitate a direct comparison, let us also  connect with the notation  of \cite{GG}, \cite{GG2}; which  is  achieved  via simple charge  redefinitions. Firstly we note that  the   4D/5D  transformations - eqs.(\ref{3.2}) and (\ref{3.7}) -  match their counterparts in \cite{GG}, \cite{GG2} after the following two redefinitions :  ${q_0}_{(5D)} \longrightarrow   ( {q_0}_{(5D)} +  p^A_{(4D)}  {q_A}_{(4D)}   )/2$  and  $p^A_{(4D)}   \longrightarrow  \sqrt{2} p^A_{(4D)}$.  We have already seen how the latter conformed to 5D split-charges and played a role in matching eq.(\ref{4.2})  to the above literature.  Now coming to the multi-ring angular momentum  in eq.(\ref{4.23}), it can be seen after some algebra that the first of the above two redefinitions simply gives a factor of 2 to the last term of eq.(\ref{4.23}). Then making  use of the second redefinition in the form  ${p^A_i}_{(5D)}   \longrightarrow  \sqrt{2} {p^A_i}_{(5D)}$ \;  produces
\begin{eqnarray}
J_{\psi} \, =  \, -  \frac{6 \sqrt{2} \pi}{G}  \sum_{i = 1}^n  L_i  {{p}^A_i}_{(5D)}  v_A  &+&    \frac{\sqrt{2} \pi}{G}  \left[  3 \sum_{i, j = 1}^n  {{p}^A_i}_{(5D)}   \left(  \widetilde{Q}_{A_{ i \,(5D) } }  -  C_{ABC}  {{p}^B_j}_{(5D)}  {{p}^C_j}_{(5D)}  \right)  \right.  \nonumber \\     &\,&  \qquad  +   \left.  2  \sum_{i, j, k  = 1}^n   C_{ABC}  {{p}^A_i}_{(5D)}   {{p}^B_j}_{(5D)}  {{p}^C_k}_{(5D)}  \right]
\label{4.33}
\end{eqnarray} 
which exactly agrees ( upto an overall factor which we leave to one's taste ) with   \cite{GG}, \cite{GG2} as  the total angular momentum of concentric black rings in ${\mathbb R}^4$.

Finally let us remark  that  writing the 5D charge   ${q_0}_{(5D)}$  in  the   form 
\begin{eqnarray}
{q_0}_{(5D)}  =  \sum_{i = 1}^n  {{q_0}_i}_{(5D)}
\label{4.34}
\end{eqnarray}  
its fragments can be easily read-off from eq.(\ref{4.23}) above. Just as was the case  earlier  with the ${{Q_A}_i}_{(5D)}$ charge, we see again that the multi-ring 4D/5D transformations for ${{q_0}_i}_{(5D)}$  are more complicated due to the presence of cross-terms which must be  carefully taken into account while performing a 4D/5D lift. In the next section, we proceed to discuss the physical  origin  of these  cross-terms and their geometric interpretation.

\section{Geometric interpretation using split-spectral flows}
In this section we try to provide  a geometric understanding of multi-black rings, based on successive application of spectral flow transformations. Such split-spectral flows  now assume  relevance  in the presence of  multiple $AdS_3 \times S^2$ horizons.  This generalises  the spectral flow discussions  of  \cite{BCDMV}, \cite{KL}  to a  multi-center setting.   

Let us first consider a single black ring, whose  near-horizon geometry is $AdS_3 \times S^2$. This will be seen to fit  exactly within  the considerations of      \cite{BCDMV}, \cite{KL}.   In this background geometry,  the 5D supergravity action  contains a Chern-Simons term 
\begin{eqnarray} 
S_{CS} = \int_{AdS_3 \times S^2}  D_{ABC} A^A \wedge F^B \wedge F^C
\label{5.1}
\end{eqnarray}  
which is not  invariant under large gauge transformations. $F^A = dA^A$ is the usual  two-form $U(1)$  magnetic flux passing through the $S^2$.  The electric charge is obtained by varying the 5D action with respect to the field strength $F^A$. Due to the presence of the  above-mentioned Chern-Simons term, the electric charge so obtained also varies  under large gauge transformations
\begin{eqnarray} 
q_A = \int_{S^2 \times S^1} \bigg( \star F_A \; + \;  3 D_{ABC} A^B \wedge F^C \bigg)
\label{5.2}
\end{eqnarray}  
Since the 5D supergravity action can be obtained from a Calabi-Yau compactification of  M-theory, the electric charge $q_A$ is the M2-brane charge from a  11-dimensional perspective ( or D2 charge in Type II A ) and the magnetic charge $p^A$ defined as 
\begin{eqnarray} 
p^A = \int_{S^2}  F^A
\label{5.2a}
\end{eqnarray} 
is the M5-brane charge\footnote{ Strictly speaking, this definition remains  valid so long as the NUT charge ( the KK monopole at the origin ) is not encompassed by the $S^2$.   } ( or D4 in 10 dimensions ).   

It can be seen by inspection that the second term in the integrand in eq.(\ref{5.2}) will decay rapidly  when evaluated  over a  homologous 3-surface sufficiently distant from the horizon, leaving only the first term to contribute. However, prior to integration,  let us consider the effect of a  large gauge transformation of $A^A$, of the type  
\begin{eqnarray} 
A^A \; \longrightarrow \;  A^A \; + \; k^A \; d \, ( \psi / 2 \pi )
\label{5.3}
\end{eqnarray} 
with $k^A$ an integer and $0 \leq \psi \leq 2 \pi$ a coordinate running along the $S^1$.  This leaves us with $A^A$-independent terms that do not vanish at infinity, thereby  producing  shifts in the electric charge $q_A$ of the type
\begin{eqnarray} 
q_A  \; \longrightarrow  \; q_A \; + \; 3 D_{ABC}  k^B p^C 
\label{5.4}
\end{eqnarray} 
This charge is clearly not gauge invariant and the physical explanation shall soon follow.  For now, let us note that this equation compares to the 4D/5D charge  transformation that we encountered earlier in eq.(\ref{3.2}), since it is the  lack of gauge invariance of the 5D  Chern-Simons term in the  action that is responsible for inducing shifts in the original gauge invariant 4D charges. 

Similarly the M-theory  angular momentum $q_0$ ( or  D0 charge  in Type II A )    is again not a gauge invariant quantity and we now proceed to derive its charge shifts, obtained via gauge transforming an  integral representation of  angular momentum.  For  a 5D supergravity action ( to be thought of as a semi-classical reduction of M-theory  in our context ), such an integral   can be extracted from appropriate contributions to the gauge field energy-momentum tensor.  For the aforesaid 5D action,  this has been derived in \cite{HOT} making  use of  Wald's method  \cite{W}  
\begin{eqnarray} 
q_0 = - \int_{S^2 \times S^1} \bigg(  \star d \xi \; + \; \star ( \xi \cdot A^A  ) \, F_A  \; + \;   D_{ABC} \, ( \xi \cdot A^A  ) \,  A^B \wedge F^C \bigg)
\label{5.4a}
\end{eqnarray} 
 Here  $\xi$ denotes the axial Killing vector with respect to the $\psi$-direction, while   $( \xi \cdot A^A  )$  is an interior product between a vector field and a one-form.  The Killing field $\xi$ generates  isometries along the $\psi$-direction; leading to a conserved charge, which is the angular momentum. In fact, the right-hand side of eq.(\ref{5.4a})  is simply the Noether charge of Wald.    Asymptotically, the $A^A$ dependent terms in the integrand ( in eq.(\ref{5.4a})  ) drop off and the integral reduces to   Komar's formula for the angular momentum. However, large gauge transforming with eq.(\ref{5.3}) yields precisely two more asymptotically   non-vanishing  remnants.  Recognizing the asymptotic form of eq.(\ref{5.2}) and  eq.(\ref{5.2a}) leads to the following charge shifts     in  angular momentum 
\begin{eqnarray} 
q_0  \; \longrightarrow \;  q_0 \; - \; k^A q_A \;  - \;  D_{ABC}  k^A k^B p^C 
\label{5.5}
\end{eqnarray} 
This again can be compared to  the 4D/5D   transformation  in eq.(\ref{3.7}). 

Now eqs.(\ref{5.4}), (\ref{5.5})   are  in fact  the spectral flow transformations in question.  The name spectral flow arises due to the fact that in the dual $( 0, 4 )$  SCFT these transformations correspond to automorphisms  of the conformal algebra.   Moreover spectral flow is a symmetry of the theory as it leaves the   generalised elliptic genus of the CFT invariant.  Note that such flows are characteristic of an odd dimensional theory. For a 4D black hole with $AdS_2 \times S^2$ horizon, the supergravity action is gauge invariant. Therefore the electric charge equals the actual number of D2 branes wrapped on Calabi-Yau 2-cycles;  while the D0 charge counts the physical D0 branes.  Because of this we can also interpret  eqs.(\ref{5.4}), (\ref{5.5})    as a 5D lift of 4D charges. 

The gauge transformation in eq.(\ref{5.3})    is picked up upon going around ( perpendicular to the $\psi$-direction  ) the ring with a probe particle; which has been given the interpretation  of  M5-anti-M5 branes being pair-produced, going around the ring in opposite directions and mutually annihilating  ( see  fig. 1  in  \cite{BCDMV}    ).   More precisely, this can  be visualised as follows.  The spatial near-horizon geometry  of a bound state of M5-M2 branes ( with angular momentum )  is  a product of Euclidean $AdS_2$  and $S^2$ ( refer to fig. 5.1 (a) below   ).   On the $AdS_2$  disc, the black ring is depicted as a circle along the $\psi$-direction. The radial coordinate  on the disc is the same as the  Taub-NUT radial direction.  Now consider the pair-production of  $k^A$ M5-anti-M5 pairs. These wrap  4-cycles on the Calabi-Yau, while the fifth direction goes around the equator of an $S^2$. This $S^2$ is a point on the $AdS_2$ disc, located on the inside of the circle representing the black ring.  The M5-anti-M5 rings along the $S^2$ equator move apart in opposite directions towards the poles, where they self-annihilate leaving behind a Dirac surface  on the $S^2$.  Since the location of the Dirac surface is unphysical, it  can  be moved away to spatial infinity. Upon  crossing  the ring, it causes a shift of   gauge potential      by an amount   $k^A \; d \, ( \psi / 2 \pi )$.    Thus the presence of a magnetic flux $k^A$ shifts the guage potential $A^A$ and consequently the charges $q_A$, $q_0$.  For the case of the  single ring described above, this flux is the dipole flux passing through the ring and is generated by its own M5 charges. Hence $k^A = p^A$ here, which leads to  eqs.(\ref{5.4}), (\ref{5.5}).

\begin{figure} [ht]
\centering
\psfrag{(a)}{$(a)$}
\psfrag{AdS_2}{$AdS_2$}
\psfrag{Black Ring}{Black Ring}
\psfrag{M5}{$M5$}
\psfrag{barM5}{$\overline{M5}$}
\psfrag{S^2}{$S^2$}
\psfrag{(b)}{$(b)$}
\psfrag{bAdS_2}{$AdS_2$}
\psfrag{Multi-Rings}{Multi-Rings}
\psfrag{1M5}{$M5$}
\psfrag{bar1M5}{$\overline{M5}$}
\psfrag{b1S^2}{$S^2$}
\psfrag{2M5}{$M5$}
\psfrag{bar2M5}{$\overline{M5}$}
\psfrag{b2S^2}{$S^2$}
\includegraphics[totalheight=0.4\textheight]{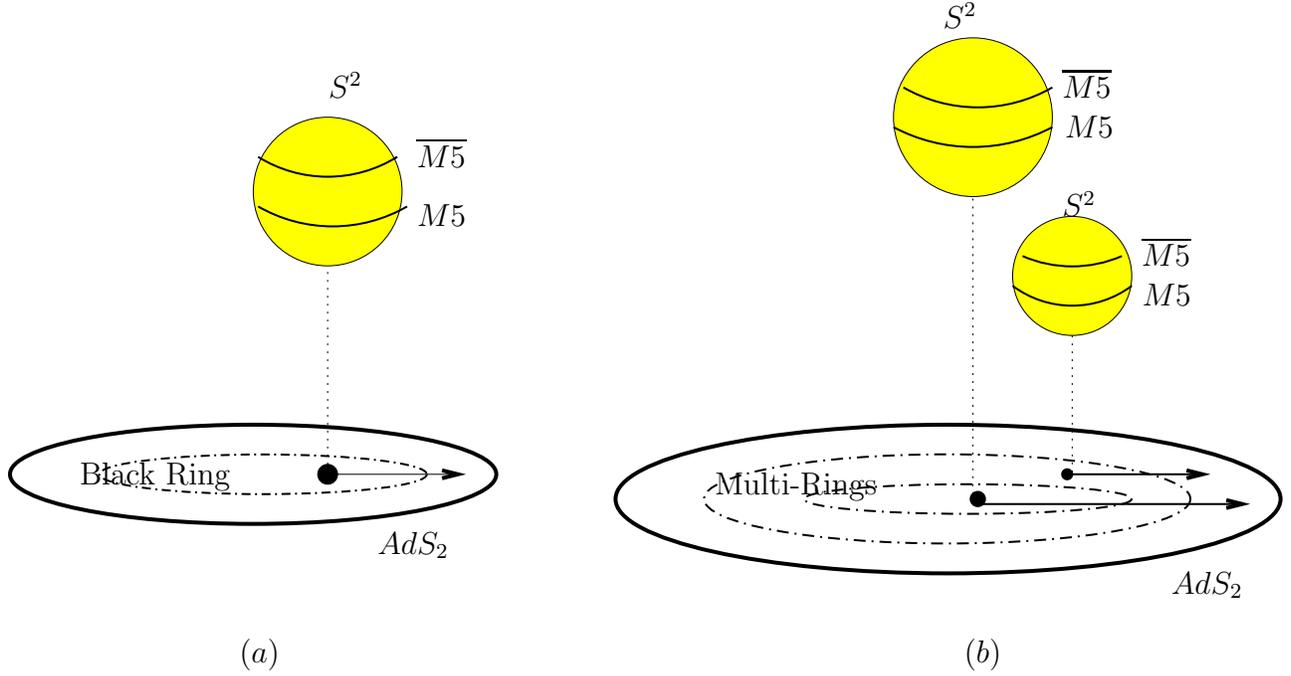}
\caption{Visualising the spectral flow for black rings : (a) Nucleation of an M5-anti-M5  pair around a single black ring leading to a large gauge transformation.   (b)  The same idea now extended to a multi-black ring background leads to multiple gauge transformations in a geometrically ordered way.  }
\end{figure}

\subsection{Electric charges and  split-spectral flows}
Now extending the above discussion,  we shall systematically derive multi-black ring  electric charges and angular momenta as a split-spectral flow argument.  We  begin with  electric charges.  Let us  label the $n$ rings with an index $i$, in increasing order of radius.  The innermost ring is labeled  $i = 1$. Its brane charges are  $p^A_1$, $q_{A_1}$, $q_{0_1}$.  Here $p^A_1$  exhibits a dipole behaviour,  generating a magnetic flux  $k^A = p^A_1$.  This in turn shifts $q_{A_1}$ by  spectral flow as  in eq.(\ref{5.4}).   Indeed this innermost  ring behaves just like the single  ring  case encountered in the previous  discussion  above.  Moving onto the next ring, this has brane charges  $p^A_2$, $q_{A_2}$, $q_{0_2}$.  As depicted in fig. 5.2 below,   the total flux passing through this ring is not only that generated by its own charge $p^A_2$,  but also that emanating from  the inner ring. These distinct fluxes give rise to the following spectral flows :
\begin{eqnarray} 
\xymatrix{  &  { \begin{array}{c} \delta = 2, \; \gamma = 2  \\  \mbox{with} \; \; k^B = p^B_2 \end{array} }  \\  
q_{A_2}  \; \; \longrightarrow \; \; q_{A_2} \; + \; 3 \; D_{ABC} \;  \; p^B_{ \delta} \; \; \; p^C_{ \gamma} \; \; \;   \ar[ur] \ar[r]  \ar[dr]  &   { \begin{array}{c}  \delta = 1, \; \gamma = 2  \\  \mbox{with} \; \;  k^B = p^B_1  \end{array}  }  \\   
  &  { \begin{array}{c}  \delta = 2, \; \gamma = 1  \\  \mbox{with} \; \;  k^C = p^C_1  \end{array}  }      }
\end{eqnarray} 
where the last transformation  occurs  due to the fact that the flux has also to be symmetrised with respect to the cycles.     The physical electric charge of this ring is then obtained by adding up all these shifts  to the original brane charge.

\begin{figure} [ht]
\centering
\psfrag{Taub-NUT}{Taub-NUT}
\psfrag{Black Rings}{Black Rings}
\includegraphics[totalheight=0.3\textheight]{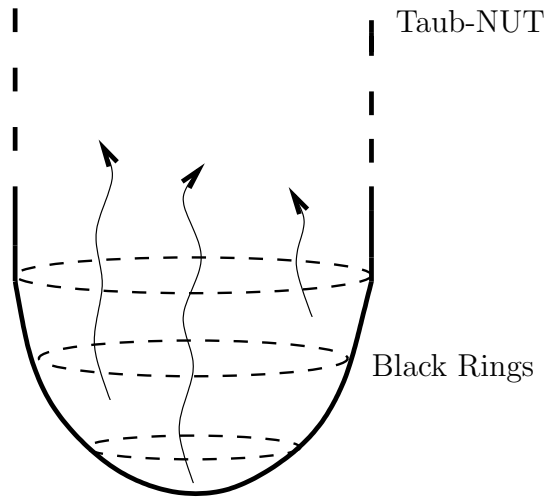}
\caption{A  Taub-NUT perspective of the  influence of  magnetic flux generated by individual black rings upon  neighbouring  black rings. }
\end{figure}

From the point of view of fig. 5.1 (b),  multi-rings are depicted as $n$-circles on the disc, one inside the other.  Nucleation of an M5-anti-M5 pair now occurs in the vicinity of each of the $n$ rings, creating $n$ Dirac surfaces. Upon moving  these surfaces  to infinity,  the  $i^{th}$-ring  is crossed by $i$ Dirac surfaces each with flux  $p^A_j$, giving in total a   flux    $k^A_{tot} =  \sum_{j = 1}^i p^A_j$  passing through this ring. This is the origin of multiple spectral flows   for  a multi-ring system.

We can now directly write  down the result   for the $i^{th}$-ring  with all the spectral flows put together :  those resulting from the intrinsic ( due to ring's own magnetic charge ) flux as  well as those from  background (  generated by those rings which are encircled by the  $i^{th}$  one  )  flux, we get  
\begin{eqnarray} 
q_{A_i}  \longrightarrow q_{A_i} + 3 D_{ABC} p^B_i p^C_i   + 3 D_{ABC}  \sum_{j =1}^{i - 1} ( p^B_i p^C_j + p^B_j p^C_i ) 
\label{5.6}
\end{eqnarray} 
  Much like the analogy in electrostatics, the fluxes due to rings which encircle the  $i^{th}$-ring  from the outside, do not affect it.  With respect to fig. 5.1 (b), each ring acts as a source, emanating  flux; while the sink is at infinity. Hence only those rings placed to the exterior of the source ring will lie  in its flux field.  Eq.(\ref{5.6}) gives us the physical charge of the $i^{th}$-ring from a spectral flow analysis. This can be compared to eq.(\ref{4.4}), where the same quantity emerged from a fragmentation analysis.

      Furthermore upon adding up  the split-spectral flow shifts of  all of the  $n$ rings  leads to the total spectral flow shift of the full  multi-ring  configuration  
\begin{eqnarray} 
Q_A^{total} &\equiv&  \sum_{i =1}^{n} q_{A_i} + 3 D_{ABC} \sum_{i =1}^{n}  p^B_i p^C_i  + 3 D_{ABC} \sum_{i =1}^{n}  \sum_{j =1}^{i - 1} ( p^B_i p^C_j + p^B_j p^C_i )  \nonumber   \\    &=&   \sum_{i =1}^{n} q_{A_i} + 3 D_{ABC} \sum_{i, j   = 1}^{n}    p^B_i p^C_j 
\label{5.7}
\end{eqnarray} 
where in the last step, the identity 
\begin{eqnarray} 
\sum_{i = 1}^{n}  \sum_{j = 1}^{i - 1}  \left( A_{ij} + A_{ji} \right)  \quad = \quad   \sum_{i = 1}^{n}  \sum_{j = 1}^{n}  A_{ij}  \; \;  -  \; \;  \sum_{i = 1}^{n}  A_{ii}
\label{5.8}
\end{eqnarray} 
has been used. Indeed $Q_A^{total}$  exactly equates to  ${Q_A}_{(5D)}$ in eq.(\ref{4.2a}),  which is simply the electric charge of a single black-ring system.   Therefore,   adding up all the spectral flow shifts as well as  the total brane charge gets us  back to the geometry  of a single black ring.  In this sense the spectral flow transforms of a multi-ring system are really split-spectral flows of a single ring system.

\subsection{Angular momenta and split-spectral flows}
The multi-ring angular momentum can now be obtained in a similar fashion.  Once again consider the $i^{th}$-ring with brane charges  $p^A_i$, $q_{A_i}$, $q_{0_i}$.  The relevant  angular momentum spectral flows for this ring are
\begin{eqnarray} 
q_{0_i}  &\longrightarrow&  q_{0_i} -  p^A_i  \sum_{j = 1}^{i }  q_{A_j}  -  D_{ABC}  p^A_i p^B_i p^C_i    \label{5.9}   \\
q_{0_i}  &\longrightarrow&  q_{0_i} -    \sum_{j = 1}^{i - 1 } p^A_j  q_{A_i}  -  D_{ABC}   \sum_{j = 1}^{i - 1 }   p^{\big( A}_j \;  \sum_{k = 1}^{i - 1 }  p^B_k \; \;  p^{C \big)}_i    \label{5.10}
\end{eqnarray} 
In the above  flow equations,  firstly we have  the intrinsic magnetic flux  $k^A_i = p^A_i$,  generated by M5 charges on the  $i^{th}$-ring itself.     This flux interacts with M2 charges as well as M5 charges ( carried on other Calabi-Yau cycles   ), both on the $i^{th}$-ring.  Then there is the  background magnetic flux $k^A_{back} =  \sum_{j = 1}^{i - 1} p^A_j$  because   this ring is placed in the background fields generated by the $i - 1$ rings to its interior.  Now a  new addition to the above is a   background  electric flux  $\sum_{j = 1}^{i - 1} q_{A_j}$, which  also  interacts with electric charges on the $i^{th}$-ring.   That explains the second term on the right-hand side of eq.(\ref{5.9}).    And   eq.(\ref{5.10})   then accounts for  interactions of  the magnetic background with the $i^{th}$-brane charges in the usual way.  The last term there has to be symmetrised and therefore the brackets in superscripts  denote  a sum over all  symmetric  permutations of   cycles.  Then   adding up all these contributions   will result in the angular momentum of the $i^{th}$-ring.  

To get the total angular momentum of the multi-ring system we add up those of each of the rings
\begin{eqnarray} 
J^{total} &\equiv&  \sum_{i =1}^{n} q_{0_i} -  \sum_{i =1}^{n}  \sum_{j =1}^{i - 1} ( p^A_i q_{A_j}  +  p^A_j q_{A_i}  )  -  \sum_{i =1}^{n}  p^A_i q_{A_i}  -  D_{ABC}  \sum_{i =1}^{n}   p^A_i p^B_i p^C_i  \nonumber   \\    &-&  D_{ABC}  \sum_{i =1}^{n}  \;  \sum_{j = 1}^{i - 1 }   p^{\big( A}_j \;  \sum_{k = 1}^{i - 1 }  p^B_k  \; \; \;  p^{C \big)}_i       \nonumber   \\              &=&   \sum_{i =1}^{n} q_{0_i} -  \sum_{i, j  =1}^{n}    p^A_i q_{A_j}  -  D_{ABC}  \sum_{i, j, k  =1}^{n}     p^A_i p^B_j p^C_k
\label{5.11}
\end{eqnarray} 
Upon  substituting $q_{A_j}$ in the last equality above  with  $\widetilde{Q}_{A_{ i \,(5D) } }$   via  eq.(\ref{4.12}),   we see that eq.(\ref{5.11})   indeed  compares\footnote{ Of course spectral flow does not determine  $q_{0_i}$ as a function of $L_i$.  That input still relies  on the integrability conditions.   }   to eq.(\ref{4.23})  leading to    $J^{total} \; = \; - \frac{G}{3 \pi} J_{\psi}$.  A split-spectral flow analysis thus provides us with a physical  understanding of where all the   different   multi-ring angular momentum  contributions actually come from.   In particular, it gives a clear  description  of  how individual rings behave in the background of other rings.

Consequently  a geometric picture of this  multi-black ring configuration  emerges from such   split-spectral flow  considerations.  In fact what these  split-flows are really doing is to break up the global multi-ring geometry into patches with locally defined gauge potentials; such that gauge fields in neighbouring patches are related upto  large  gauge transformations. In  fig. 5.1 (b)  these patches can be  identified  as follows : first there's the innermost disc inside the first ring, defining a patch with gauge potential  $A^A_1$;  then there are the  annular regions all around it, with gauge potentials    $A^A_2$, $A^A_3$,.........       respectively.    This defines a chain of  potentials spanning the entire geometry 
\begin{eqnarray} 
A_1  \;\; \mathop{\longrightarrow}\limits^{\beta_1} \;\;  A_2 \; \; \mathop{\longrightarrow}\limits^{\beta_2}\; \; A_3 \; \; \cdots  \cdots   \cdots  \cdots  \; \; \mathop{\longrightarrow}\limits^{\beta_{n - 1} }  \;  \; A_n  \; \; \mathop{\longrightarrow}\limits^{\beta_n}\; \; A_n + \beta_n 
\label{5.12}
\end{eqnarray} 
( suppressed vector indices may be readily reinstated here ) the  $\beta_i$  are large gauge transformations between $A_i$  and  $A_{i + 1}$.   In fact these local regions emerging here due to split-spectral flow considerations might provide a conceptual basis for the analysis of \cite{HOT}  where the authors  compute localised charge integrals for black rings by dividing  the geometry into local patches which are all glued together. The existence of such  patches enable near-horizon integrals such as those in eqs.(\ref{4.7}), (\ref{4.8})  to capture all the data normally extracted from  the full geometry.

\section{Conclusions and outlook}
Two remarkable set of ideas pertaining to string theoretic descriptions of  black holes, that have generated   lots  of  excitement  in the aftermath of the OSV conjecture are : (1) the 4D/5D connection between black holes/rings \cite{GSY}, \cite{GSY2}; and (2) multi-center black holes as non-perturbative corrections to the black hole partition function \cite{DGOV}.  In this note we have sought for  a  modest  attempt  at combining these two,  in the sense of the commutative box diagram of  eq.(\ref{1.3}).  

We approach the problem by setting-up an explicit 5D construction of black ring fragmentation and thereafter also show that fragmented black rings are equivalent  to a direct 5D lift of 4D multi-black holes.  For the purposes of the latter, we determine the multi-center 4D/5D charge transformations as well.

Related to these events is the important issue of interpretation of charges in 5D, especially for our multi-center  split charges. In \cite{HOT} it was shown that the electric charge  ( and angular momentum  ) of a  single black ring could be expressed purely in terms of near-horizon data as a Page charge. In our analysis we see that the  5D charges ${{ {Q}_A}_i}_{(5D)}$  which participate in  fragmentation  are in fact also Page charges \; ( \;  as opposed to being Maxwell  charges \; ) \;  and in that sense these  are the physical charges of the system.   Whereas the multi-center charges  $\widetilde{Q}_{A_{ i \,(5D) } }$  that usually  appear in the supergravity multi-ring metric  are not physical  charges. Even though the latter-mentioned  charges can be algebraically  related to the former ones, we find it  nevertheless important to distinguish  the physically relevant ones  for the  multi-ring configuration.

A rather interesting application of the 5D fragmentation methods developed in this  note is an alternative derivation of the angular momenta of concentric black rings. It is indeed gratifying to note that we are able to exactly reproduce the results of Gauntlett and Gutowski.

Lastly, we saw  how the  introduction of  split-spectral flows lends a geometric perspective to  shifts in brane charges of fragmented black rings by accounting how a Dirac string generated by a given ring influences other rings in such a multi-ring background.  This serves as yet another derivation for the total angular momentum of a multi-ring system. Moreover summing up all the split-spectral flow shifted charges of all the fragmented rings exactly gives back the observed electric charge of a single black ring.  The split-spectral flows basically divide the geometry into patches with locally defined gauge fields. The significance of these patches becomes relevant when computing near-horizon integrals.

From a broader perspective, one might contemplate over the role of fragmented configurations on the black hole/ring partition function. In \cite{DGOV}, each fragmented configuration is  viewed as a multi-AdS throat geometry; and further following \cite{Br}, \cite{MMS}, each such geometry is associated to some  saddle point of the  partition function. In that sense $Z_{BH}$ is presumed to  sum over all possible geometries subject to charge  conservation constraints. Fragmentation is thus a euclidean tunneling process from one minima to another.  These leading order contributions therefore  dominate the  multi-AdS partition sum  of  \cite{DGOV}.   However there ought to be further sub-leading corrections to each multi-center configuration that  should be computable from  any  complete partition sum. At this stage, it would be very tempting to think that the black hole farey tail partition function of \cite{DMMV}, \cite{BCDMV}, \cite{KL} might be precisely the object that captures the multi-center saddle points as well as its  sub-leading  corrections. Whether or not these multi-center geometries lend a physical description to the farey tail story remains to be seen.

\section*{Acknowledgements}
Foremost I would like to thank Erik Verlinde for lending his guidance, patience and insight into this project. Others from whom I also  enjoyed learning  by way of discussions include Dumitru Astefanesei, Miranda Cheng, Jan Manschot   and   Ilies Messamah.  For help with the figures and latex doubts, I'd like to acknowledge Michele Maio and  Meindert  van der Meulen respectively. This research is financially supported by De Stichting voor Fundamenteel Onderzoek der Materie (FOM).

\end{document}